\documentclass[12pt]{article}
\usepackage{epsfig,amssymb}

\newcommand{\bea}{\begin{eqnarray}}
\newcommand{\eea}{\end{eqnarray}}
\newcommand{\beq}{\begin{equation}}
\newcommand{\eeq}{\end{equation}}

\newcommand{\bi}{\begin{itemize}}
\newcommand{\ei}{\end{itemize}}

\newcommand{\ua}{\uparrow}
\newcommand{\da}{\downarrow}

\newcommand{\bc}{\begin{center}}
\newcommand{\ec}{\end{center}}
\newcommand{\sqpi}{{\sqrt{\pi}}}

\newcommand{\on}{{\bar\omega_n}}

\input epsf.sty

\begin{document}

\begin{center}
{\bf \Large Transport in Quantum Wires}
\end{center}
\vskip .5 true cm
\centerline{\bf Siddhartha Lal$^1$, Sumathi Rao$^2$  and Diptiman Sen$^1$} 
\vskip .5 true cm

\centerline{\it $^1$ Centre for Theoretical Studies,}  
\centerline{\it Indian Institute of Science, Bangalore 560012, India} 
\vskip .5 true cm

\centerline{\it $^2$ Harish-chandra  Research Institute,}
\centerline{\it Chhatnag Road, Jhusi, Allahabad 211 019, India}
\vskip .5 true cm

\begin{center}
{\bf Talk presented\footnote{by Sumathi Rao} at the \\
``International Discussion Meeting on Mesoscopic \\ and  Disordered
Systems''}
\end{center}
\vskip 1.6 true cm

\begin{abstract}
With a brief introduction to one-dimensional channels and conductance
quantisation in mesoscopic systems, we discuss some recent 
experimental puzzles in these systems, which include reduction of
quantised conductances and an  interesting {\it odd-even} effect
in the presence of an
in-plane  magnetic field. We then discuss a recent 
non-homogeneous Luttinger liquid model 
proposed by us, which addresses and gives an 
explanation for the reduced conductances and the 
{\it odd-even} effect.
We end with a brief summary and discussion of future projects.

\end{abstract}

\section{Introduction} 

Recent advances in the fabrication of gating of the two-dimensional
electron gases formed at the inversion layer of high mobility 
GaAs-AlGaAs heterostructures, have enabled the experimental 
study [1-7] of electron
transport through very few channels or even a single channel, not
all of which has been theoretically well-understood. 
This has caused a tremendous
upsurge in the study of quantum wires and Luttinger liquids 
(LLs) \cite{review}.

Early experiments \cite{earlyexpts} found that the dc conductance through
narrow constrictions were quantised with the steps separated by 
$\Delta g = {\tilde g}~(2e^2/h) ~$, where ${\tilde g}\sim 1$. 
This could be understood within the Landauer-Buttiker picture
of transport through the scattering matrix approach \cite{datta}.
In fact, when electron-electron interactions were included via
the LL theory, initial calculations \cite{kane} expected a reduction in 
quantised conductance - $\tilde g$ was expected to be the Luttinger parameter,
which depended on the strength of the interactions. Later it was
realised \cite{leads}
 that the Fermi liquid leads play a role, and when that
was taken properly into account, $\tilde g\sim 1$, even for a LL wire.

However, in the last few years, it has been found in several 
experiments [2-7]
that $\tilde g<1$ and in fact, varies, depending on the temperature $T$ and
the length of the wire $L$. However, conductance quantisation or
plateaux were still seen, indicating that the reduction was not due
to impurities, and moreover, the reduction was found to be uniform
in all the channels. Also, 
the flatness of the plateaus appeared to indicate an insensitivity to the 
electron density in the channel. Besides this, when an 
external 
magnetic field was placed in-plane and parallel to the channels \cite{liang},
a splitting of the conductance steps 
was  observed together with an  odd-even effect of 
the renormalization of 
the plateau heights, with the odd and even plateau heights being 
renormalized by smaller and larger amounts respectively.
This motivated us to study a LL model of a quantum wire,
with Fermi liquid leads and the unusual feature of additional
lengths of short LLs modeling the contacts between the leads
and the wire.

The plan of the paper is as follows. In Sec.II, we review conductance 
quantisation of one-dimensional channels in mesoscopic systems. In
Sec.III, we give a brief description of some recent experimental 
results and some of the puzzles mentioned above,
such as reduction of quantised conductance even in clean quantum
wires and 
a novel  odd-even effect in the presence of magnetic field.
In Sec.IV, we give a brief introduction to Luttinger liquids (LLs) and
bosonisation and transport in these models. 
 In Sec.V, we describe  
our model, which is essentially an inhomogeneous LL model, with
five distinct regions, the two {\it leads} and {\it  contacts} 
and the central {\it wire} and having two barriers at the boundaries
between the contacts and the leads and explain how this model is 
motivated by the experimental results. 
In Sec.VI, we describe  the results of our
calculations with the model, showing reduced conductances as a
function of temperature, and lengths of the wire and the contacts.
In Sec.VII, we show that in the presence of an in-plane magnetic
field, the barriers show different renormalisations for the odd and 
even plateux leading to the odd-even effect.
Finally, we end in Sec.VIII, with 
a brief summary and possible future extensions of our work.  

\section{ Introduction to 
one-dimensional channels and  conductance quantisation}

Semiconductor mesoscopic conductors are fabricated by confining electrons
to a two-dimensional conducting layer formed at the interface
between GaAs and GaAlAs. The electrons are free to move in the
$x-y$ plane, but are confined by some potential in the $z$-direction. 
Their dispersion is given by $E=E_n + {\hbar^2\over 2m}(k_x^2
+k_y^2)$.
The index $n$ labels different sub-bands. At 
low temperatures and low carrier densities, only the 
lowest sub-band $n=0$ is occupied. So we can ignore the $z$-dimension
altogether and treat the electron gas as a two-dimensional electron
gas (2DEG) in the $x-y$ plane.

Now, we apply strong confining potential in  the $y$-direction. If
the dimensions of the conductor were large, then the conductivity of
the sample would be given by $g=\sigma W/L$, where $\sigma$ is the
conductivity of the material and $W$ and $L$ are the widths and
lengths of the sample. But as the length is reduced, experimentally,
it was  found that the conductance reaches a limiting value, $g_c$
instead of increasing indefinitely. The reason for this is  that there
exists a non-zero resistance at the interface between the sample and
the leads, which is unavoidable. This was called the contact
resistance and for very clean samples, at low temperatures, it was
found to be a universal quantity.

This can be very easily understood, using Landauer's
picture \cite{datta} of electron transport as a scattering problem.
Consider applying a strong confining potential in the $y$-direction,
so that the dispersion is given by 
$E=E_0+\epsilon_m +{\hbar^2k_x^2/2m}$.
For each value of $m$, we have a different sub-band.
The spacing between sub-bands is fixed by the confining potential; the
stronger the
confinement, the  farther apart the sub-bands.
In actual experiments, confinement in the $y$ direction is controlled by a
gate voltage. We have a 
quasi-one-dimensional  wire when only a few sub-bands are occupied.

Let us now calculate the conductance through each sub-band or channel
of the  wire for an  applied voltage bias
$\mu_2-\mu_1$.
The current $I$ is given by 
$I=(1/L) \sum_k e v = (e/ L) \sum_k 
(\partial E/ \partial \hbar k)$
because the density of electrons for each $k$ state is $1/L$.
Since $\sum_k = 2 \times (L/2\pi) \int dk$, we get 
\beq
I={2e\over h} \int_{\mu_1}^{\mu_2} dE = {2e^2\over
h}{(\mu_2-\mu_1)\over e}~.
\eeq
Hence, the minimum contact resistance per channel is given by
${h/2e^2}$.
For $M$ channels, the same argument gives 
$I=(2e^2/
h)M((\mu_2-\mu_1)/ e)$, which implies that  
the contact resistance = ${h/ 2e^2M}$. Since a macroscopic
conductor has  a very large number of channels, this explains why
for macroscopic conductors, the contact resistance
is negligible.
But for a single channel, $R_c = 12.9 k\Omega$, and is certainly not
negligible.

\begin{figure}[h]
\begin{center}
\leavevmode
\epsfysize=5truecm \vbox{\epsfbox{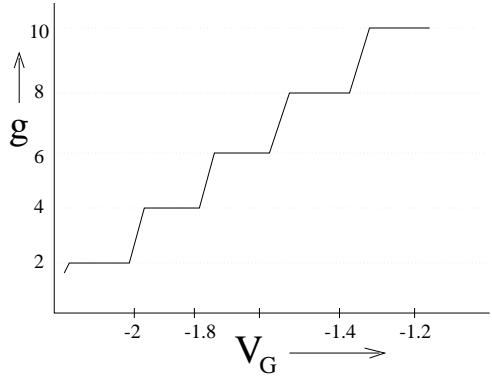}}
\end{center}
\caption{Quantised conductance of a ballistic wire. A negative
gate voltage is applied to deplete the electrons in a narrow
channel. The conductance $g$ is measure in units of $e^2/h$.}
\label{nk1} 
\end{figure}

This contact resistance has been experimentally
measured \cite{earlyexpts}.
A gate voltage controls the density of electrons in the channel
and narrows down the constriction progressively. 
The current is plotted as function of the gate voltage. Although, the
width of the constriction changes continuously, the current goes down 
in steps as seen in Fig.\ref{nk1}.
So starting with current through a single channel, the current
increases 
each time a sub-band comes below the Fermi level. When $M$ 
sub-bands are filled,
$g={2Me^2 / h} $ - $i.e.$, the steps are 
quantised at multiples of $2e^2/h \equiv g_0$. For very precise 
quantisation, one needs ultra-clean samples where the only source
of resistance is the 
contact resistance. 

This quantisation can be shown to be true even when the electrons are
interacting. That is not a surprising result, when one realises that
the  resistance that is being measured here is purely a 
contact resistance and does not depend
upon the interior of the wire.

\section {Brief description of recent experimental \\ results}

Here, we give a brief summary of the experimental  results on
transport in quasi-one-dimensional channels 
in the last few years. One of the main 
features  that has been found is the reduction of 
quantised conductances from $N \times g_0$.
Flat plateaux, independent of the gate
voltage  are still seen in each channel, but the quantisation is now
at some value below  $N \times g_0$.
(See, for instance, Figs. 2 and 3 in \cite{yacoby}.)
This has been seen by several groups \cite{tarucha,yacoby,facer,liang,thomas} 
and the variation of the reduction as a function of the temperature and
the length of the quantum wire has also been measured. Tarucha {\it et
al} \cite{tarucha} performed experiments 
with wires of lengths of $2\mu m$ to $10\mu m$ fabricated using split-gate 
methods at temperatures from $0.3K$ to $1.1K$,
and found deviations from the perfect quantization of the steps.
Yacoby {\it et al} \cite{yacoby}
made measurements on a 2$\mu m$ wire and at temperatures
ranging from $T=0.3K$ to $T=25K$ and they found that the reduction is
surprisingly uniform for each channel.
They also found that the step heights were increasingly renormalised
at lower temperatures.   Similar experiments
\cite{facer,liang,thomas}
studied the variation in
the reduction as a function of both length and temperature and
confirmed that the step heights were increasingly 
renormalized as either the temperature 
was lowered (for a fixed length of quantum wire) or the length of the 
quantum wire was increased at a fixed temperature.
Such renormalizations would require back-scattering of electrons.
If these back-scatterings were due to impurities within
the quantum wires, the conductance corrections would be gate
voltage dependent as shown in our calculations. This can
certainly not lead to flat conductance plateaus as seen in the experiments.

In the presence of a magnetic field, Liang {\it et al} 
\cite{liang} found 
that as they turned up the external magnetic field (kept 
in plane and aligned along the direction of the channel) from $0$ to $11T$, 
the the spin degeneracy gets lifted and  
each conductance step splits into two steps, 
with the heights of both being less than 
$g_0$. At a magnetic field strength of $11T$, they found that the 
difference between the conductance of successive pairs of spin-split sub-bands 
alternates. This shows that the conductance of the odd numbered 
spin-split sub-bands containing the moments aligned with the magnetic field 
undergoes little renormalization, 
while the conductance of the even numbered spin-split sub-bands containing 
the moments anti-aligned with the magnetic field undergoes a large 
renormalization. This is the odd-even effect.


The naive non-interacting Landauer explanation for the renormalisation
of the plateaux would  
require back-scattering of electrons
due to impurities within
the quantum wires, which could lead to  
$g=g_0 \times N \times T$ where $T$ is the transmission
coefficient.
But $T$ is a function of the energy $E$ of the electrons, which in
turn, is related to density of electrons in the channel, (which is
controlled by the gate voltage $V_G$). Thus, 
the conductance corrections would be gate
voltage dependent. This can
certainly not lead to flat conductance plateaus as seen in the experiments. 
To obtain flat plateau renormalisations, one would have to postulate
that $T$ is independent of $E$, which is very  unlikely. Moreover, the
conductance corrections depend on temperature. This is hard to arrange
within the usual non-interacting Landauer model of conductance
corrections. Neither is the odd-even effect easy to see within
the standard picture.


\section{Introduction to Luttinger liquids and \\ bosonisation}

The standard paradigm for many-fermion systems is Fermi liquid theory. 
The Fermi gas is a  collection of non-interacting fermions, filled
upto the Fermi level. 
Excitations over the ground state are quasiparticles ( above the
Fermi surface)  and quasiholes ( below the Fermi surface),
which have the same quantum numbers as that of the original 
electrons or holes.
The idea behind Fermi liquid (FL) theory, is that interactions can change
the ground state, modify the excitations  and their energies and so on, but
essentially, one continues to have single-particle fermion like
excitations even after inclusion of the interactions. 
These excitations (called Landau quasiparticles) 
can have their masses, couplings, etc,
renormalised, but basically each state is in one-to-one correspondence
with the non-interacting states.

In three dimensions, most electronic phenomena (with some exceptions
like the two-channel Kondo problem) can be understood
within the framework of FL theory. In two dimensions, there  exists 
some phenomena, where it is not clear
whether FL theory is really 
applicable. For instance, many people believe that high $T_c$
superconductivity needs non-FL behaviour.
Also, the $\nu =1/2$ state
in FQHE is probably an example of a non-FL.

\begin{figure}[h]
\begin{center}
\leavevmode
\epsfysize=5truecm \vbox{\epsfbox{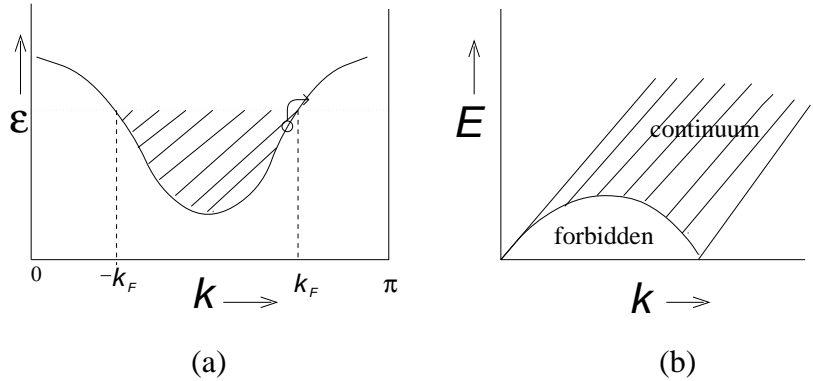}}
\end{center}
\caption{a) Single particle spectrum of the Fermi gas in
one dimension.$\quad$ b) Particle-hole pair spectrum with the forbidden
region and the continuum.}
\label{nk2} 
\end{figure}

But in one dimension, it is well-known that FL theory breaks
down, and the ground state is a non-trivial state, called the
{\it Luttinger liquid} (LL) state,  
which no longer has quasiparticles similar to that
of the non-interacting case. Why is one dimension different?
The reason is that the
Fermi surface is reduced to two points. Hence, low-energy  
particle-hole pair excitations have their energy fixed by momentum 
and vice-versa, (see Fig.\ref{nk2}) and consequently,
the particle-hole mode propagates coherently as a  new
bosonic particle. All 
low energy modes are exhausted by these bosonic modes; hence,
the fermionic theory can be rewritten in terms of bosonic fields using
a procedure called bosonisation.

The interesting point is that it is possible to 
solve a non-trivial fermion model by recasting in boson language. In
fact, even when the bosonic model is  not exactly solvable,
bosonisation turns out to be a useful tool 
to compute
correlation functions, which can then be 
supplemented with  renormalisation group (RG)
arguments  to get useful information.

\subsection{Bosonization}

The explicit relation relating bosons and fermions is given by
\beq
 \psi_R \sim {\eta_R\over \sqrt{2\pi}} e^{-2i{\sqrt{\pi}} \phi_R} 
\eeq
for a right-moving fermion $\psi_R$ and a right-moving boson $\phi_R$ and 
similarly for left-movers.
$\eta_R$ is a  Klein factor which  takes care of the 
anti-commutation property of fermions.
For any interaction, one can show that the velocities of the
spin and charge modes
are different; so generically, a LL state has spin-charge separation. 
Also, instead of a pole in the
single particle propagator, even when interactions are included, as one
would expect for a FL, here one finds that all
correlation functions have a power law fall-off, with the power being
dependent on the interaction parameter. These anomalous exponents in various
correlation functions or response functions are  the hallmark of
LL behaviour.

A generic LL model given by  
\beq
H = {v\over 2} \int dx ~[K \Pi^2 +
{1\over K}(\partial_x\phi)^2]~.
\eeq
For non-interacting fermions $K=1$ and  $v=v_F$. For 
repulsive interactions $K<1$.
All low energy properties can be found in terms of $K$ and $v$.
But one problem with LL models is that
contact with the microscopic theory is not  obvious.

\subsection{Transport and Conductances in LL}

In this subsection, we study transport, 
in particular the dc (or zero 
frequency) conductivity, in clean one-dimensional  wires, with no
impurities or barriers.

\vskip 0.3cm
\noindent {\bf Conductance of a clean LL}
\vskip 0.2cm

First, we shall perform a calculation to compute the conductance of a
LL without any consideration of contacts or leads.
By computing the conductance using the current-current correlation
functions in the Kubo formula, 
we can show that the conductance of a clean Luttinger wire
(with no leads) is given by
\beq
g\equiv {I\over V}= {Ke^2\over 2\pi} ~.
\eeq
But when we include leads modeled as semi-infinite Fermi 
liquids on either side
($i.e.$, we study the same bosonic model, but with $K$ varying
spatially -  $K=K_{L} =1$ for $x<0$ and $x>D$, where $D$ is the
length of the wire, and $K=K_W$ for $0<x<D$), then 
we find that
\beq
g = {K_{L}e^2\over 2\pi}  = {e^2\over 2\pi}~.
\eeq
This is not a surprising result. As we have explained before, the
resistance is only due to the contacts. So whether the electrons in
the wire interact or not, is not relevant to the conductance.

\vskip 0.3cm
\noindent {\bf Transport through single impurity}
\vskip 0.2cm

Here, we study the
conductance through  a single impurity.
We will see that 
interactions change the picture dramatically. For a non-interacting
one-dimensional wire, from just solving usual one-dimensional quantum 
mechanics problems, we know that we can get both transmission and reflection
depending on the strength of the scattering potential. But for an
interacting wire, we find that for any scattering potential,
however small, for repulsive interactions between the electrons, there
is zero transmission and full reflection (implies conductance is zero,
or that the wire is `cut')
and for attractive interactions between electrons (which is of course
possible only for some renormalized `effective' electrons), there is
full transmission and zero reflection (implying perfect conductance
or `healing' of the wire). {\bf Note:} The above is true for $T=0$ and for
$L\rightarrow \infty$, where $T$ is the temperature and $L$ is the
length of the wire. For finite temperature and lengths, there will be 
calculable corrections to the above results, which, in fact, is what we
shall explicitly  compute in our model.

Let us introduce a single impurity (a barrier or constriction, or even
an experimentally
introduced strong bias at a point along the wire) in the quantum wire.
We can model this 
by a potential $V(x)$ around origin and write the Hamiltonian for the
system as 
\bea
\delta H &=&  \int dx V(x) \psi^{\dagger}(x) \psi(x) 
~=~ -\lambda [\psi_{R}^{\dagger}(x=0)\psi_L(x=0) + h.c.] \nonumber\\ 
 &=& \lambda \cos 2{\sqrt\pi} K \phi (0)~. 
\eea
In the weak barrier limit, the renormalisation group (RG) equation for the 
barrier strength gives
\beq
\frac{d\lambda}{dl}=(1-K)\lambda ~.
\eeq
Thus, the perturbation is relevant 
when $K<1$, and the strength of the potential grows as we renormalise
down to lower temperatures or longer lengths. At $T=0$ or
$L\rightarrow \infty$, this implies that the wire is cut.

Since, the RG equation is perturbative in $\lambda$, it is necessary
to look at the strong barrier limit and 
ask what happens when we start from a cut wire and 
allow for hopping. 
Here, the perturbation term or the hopping term is given by
\beq
\delta H ~=~ -t [\psi_{<}^{\dagger}(x=0)\psi_>(x=0) + h.c.]   
~=~ t \cos {2{\sqrt\pi} \over K} \theta (0)~,
\eeq
and the RG equation (perturbative in $t$) is given by
\beq
\frac{dt}{dl}=(1-{1\over K})t~.
\eeq
For $K<1$,  $t \rightarrow 0 $ - $i.e.$, the hopping term, is
irrelevant and the  
the wire remains `cut'.

In either case, the 
conductance is zero.
For finite $T$ and $L$, there are temperature and length dependent
corrections. 


\section{Our one-dimensional LL model}

One might expect that 
the simplest model to obtain renormalisations of the 
quantised conductances  would be a LL model between
two FL leads, with barriers at the junctions between the
leads and the wires. However, the problem in such a model  is that the
velocity of the electron in the wire is set by 
$v_w = \sqrt{v_F^2 - 2E_s/m}$ where 
$E_s$ are the  discrete energy levels of the confinement potential set by
the gate voltage $V_G$. As $V_G$ changes, the density of electrons
in the channel varies and $E_s$ clearly depends on the channel number $s$.
These velocities are, hence, density and channel dependent. 
The interaction parameter is also expected to be
density and channel dependent. Hence, we 
cannot expect flat and channel independent renormalisations of the
conductance plateaux, using such a model.

Instead, in our model for the quantum wire, we include 
two additional contact regions  modeling the change from the 2DEG 
to the 1DEG. The contacts themselves are chosen to be short lengths
 of LL wire. Since, 
interactions are expected to be important in this region, we take
$K=K_C$. But the important point is that this 
region is independent of the gate voltage, so no 1D bands are formed and
$v_C$ is  set by $v_F$ and interactions in the contact, and is not influenced
by the gate voltage that controls the density in the wire.

In fact, Yacoby {\it et al} \cite{yacoby2}
 established the existence of $l_{2D-1D}$, which is the
length of the scattering region between the 2D leads and the 1D wire.
They found that a length  
$l_{2D-1D} \sim 2-6 \mu m$ was needed to cause backscattering.
Such a scattering  length existed in their original experiment as well
and was responsible for reduction from ideal quantised 
conductance value. 
Hence, our model that a finite contact region is needed to 
obtain the smooth transition from the 2DEG to
the 1D wire is  
physically quite reasonable.

So our model for the quantum wire, motivated both by the way the 
wire is  constructed as well as by the experimental results,
is a LL model, but in five pieces. 
The external 2DEG reservoirs are  modeled as two semi-infinite 
FL leads, while  
the contacts are modeled
as short quantum wires with the junctions at either end modeled as 
$\delta$-function barriers. The inter-electron interactions in the
system, and hence the parameter $K$ which characterizes the interactions, vary 
abruptly at each of the junctions. Hence, we study a
$K_L$-$K_C$-$K_W$-$K_C$-$K_W$ model.
The barriers at the junctions are because 
changes in geometry ( if not adiabatic) and also changes in
interaction parameters do lead to barrier-like potentials at the junctions.
The important point is that $v_C$ and $K_C$ are independent of gate voltage.
We assume weak barriers at the contact-lead junction
and 
negligible barriers at the contact-wire junction, because the 
change in geometry and interaction strengths are likely to be
stronger at the contact-wire junction.

Hence, the model can be written as
$$
S_0 = \int dt [ \int_{-\infty}^0 dx {\cal L}_1 + \int_0^d dx {\cal
L}_2 + \int_d^{l+d} dx {\cal L}_3 
+\int_{l+d}^{l+2d} dx {\cal L}_2 
+\int_{l+2d}^{\infty} 
dx {\cal L}_1 ]~,  
$$
where
\bea
{\cal L}_1 ~&=&~ {\cal L} (\phi_\rho ; K_L , v_L ) ~+~ 
{\cal L} (\phi_\sigma ; K_L , v_L ) ~, \nonumber \\
{\cal L}_2 ~&=&~ {\cal L} (\phi_\rho ; K_{C\rho} , v_{C\rho} ) ~+~ 
{\cal L} (\phi_\sigma ; K_{C\sigma} , v_{C\sigma} ) ~, \nonumber \\
{\cal L}_3 ~&=&~ {\cal L} (\phi_\rho ; K_{W\rho} , v_{W\rho} ) ~
+~ {\cal L} (\phi_\sigma ; K_{W\sigma} , v_{W\sigma} ) ~ \nonumber 
\eea
with 
\beq
{\cal L} (\phi ; K,v) = {1\over 2Kv} (\partial \phi /\partial t)^2 -
{v\over 2K}
(\partial \phi /\partial x)^2~.
\eeq 
The action for the coupling to the gate voltage, as well as the
junction-barrier terms, are given by
\bea
S_{gate} + S_{barrier} &=&  \frac{eV_G}{\sqpi} \int dt [\phi_{3\rho} 
- \phi_{2\rho} ] 
+ \frac{V}{2\pi \alpha} ~\Sigma_i \int dt [ \cos (2 \sqpi 
\phi_{1i}) \nonumber \\
&+& \cos (2 \sqpi \phi_{4i} ~+~ \eta ) ]  
\eea
where $i$ is summed over $\ua ,\da$, 
 $\alpha$ is a short distance cutoff and  
$\eta$ in terms of the wave numbers in the contact
and wire regions  is given by $\eta = 2 k_C d + k_W l$.

Once, we have the above action, there exists a straightforward method
to study it. We  
integrate out the bosonic degrees of freedom 
everywhere except at $x=0,d,l+d$ and $x=l+2d$ and 
study the effective action in various limits.
Here, we just quote the results\cite{us1,us2} that we obtain in
the various limits. 

\noindent $\bullet$ High temperature or high frequency limit 

The Matsubara frequencies $\on$ are quantized in multiples of the 
temperature as $\on = 2\pi n k_B T$ and hence temperature is 
equivalent to  frequency. When 
$ T \gg T_d \equiv  v_{C\rho}/(2\pi k_B d)$ and $T_l \equiv 
v_{W\rho}/(2\pi k_B l)$, the 
conductance correction is given by 
\beq
g ~=~ {2e^2\over h} K_L ~[~ 1 - c_1 T^{2(K_{eff} - K_L)} 
 ~(|V(0)|^2 + |V(l+2d)|^2) ~]~.  
\eeq
Here, $c_1$ is a dimensionful constant depending on 
$v_{C\rho}$ and the cutoff $\alpha$ (but  
independent of all factors depending  on gate voltage $V_G$)
and 
\beq
K_{eff} ~=~ \frac{K_L K_{C\rho}}{K_L + K_{C\rho}} ~+~ \frac{K_L 
K_{C\sigma}}{K_L + K_{C\sigma}} 
~=~ \frac{ K_{C\rho}}{1 + K_{C\rho}} ~+~ \frac{1}{2}~.  
\eeq
The only parameters are $c_1$ and the interaction parameter $K_{C\rho}$
because $K_{C\sigma}=1$ in the absence of a magnetic field.
For the Yacoby {\it et al}  experiment, $l_{2D-1D}=6\mu m$, which means
that  $T_d, T_L \sim 0.1-0.2 K$, so that all their results are
in this regime. Thus, if we assume that their $l_{2D-1D}$ is what
we are calling the contact length $d$, then, this result explains
why they get flat plateau renormalisations and renormalisations independent
of the channel, since $K_c$ and $v_C$ are independent of the 
gate voltage.
From this analysis, one
expects a similar order of magnitude for the contact length 
$l_{2D-1D}$ even in the  Tarucha {\it et al} and 
Liang {\it et al} experiments.
In fact, our results also tell us that as the 
temperature $T$ is raised,
the conductance 
corrections get smaller and  approaches 
integer multiples of $g_0$, as indeed seen in the experiments.
 
Note that our results 
crucially depends on introduction of contact region and placing of the 
barriers between contact and lead. In fact, the barriers can be anywhere 
in the contact, as long as they are not too close to the wire region, and
we still get the same result.
Essentially, by modeling the contacts as short LL's, we
have included many-body effects at the contacts 
and shown that they are
responsible for the observed conductances.

\noindent $\bullet$ Intermediate temperatures 

Here, there are two possible scenarios. Either, we have 
a) $T_l \ll T \ll T_d$ ~~~ (if $d \ll l$), or we have 
b) $T_d \ll T \ll T_l$ ~ (if $l \ll d$). 

For scenario $(a)$, which we call the wire regime,
\beq
g = {2e^2\over h} K_L [ 1 - c_2 T_d^{2(K_{eff} - {\tilde
K_{eff}})}
T^{2({\tilde K_{eff}} - K_L)}  
~(|V(0)|^2 + |V(l+2d)|^2) ~]~  
\eeq
where
\beq
\tilde K_{eff} ~=~ \frac{K_L K_{W\rho}}{K_L + K_{W\rho}} ~+~ \frac{K_L 
K_{W\sigma}}{K_L + K_{W\sigma}} ~ 
~=~ \frac{ K_{W\rho}}{1 + K_{W\rho}} ~+~ \frac{1}{2}~.  
\eeq
Here, clearly, the conductance depends on 
the wire parameters and hence on the gate voltage; in this regime, we would
not expect flat and channel independent plateau renormalisations.

Scenario (b), which we call the quantum point contact limit, however,
has channel independent flat plateau renormalisations both for the
high and intermediate temperature regimes.

\noindent $\bullet$ Low temperatures 

Here, $T \ll T_d$ and $T_l$ and in this regime, the conductance is
given by
\bea
g &=& {2e^2\over h} K_L [ 1 - c_3 T^{2(K_L - 1)}
T_d^{2(K_{eff} - {\tilde
K_{eff}})} 
T_l^{2({\tilde K_{eff}} - K_L)}   ~|V(0) + 
V(l+2d)|^2 ~]~ \nonumber  \\
 &=& {2e^2\over h}  [ 1 - c_3 
T_d^{2(K_{eff} - {\tilde
K_{eff}})}
T_l^{2({\tilde K_{eff}} - 1)} 
~|V(0) + 
V(l+2d)|^2 ~]~.
\eea

In this regime, there is 
no temperature dependence. The conductance correction  
only depends on $d$ and $l$, since they take over the role
of the infra-red cutoff.
As $d$ and $l$ decrease, corrections become smaller and $g$ approaches
$g_0$. In this case, there also exists the 
possibility of resonant transmission, since the barriers  are
seen coherently. The conductance corrections in this regime
do depends on wire parameters and hence gate voltage, and so one
would not expect flat plateau renormalisations. 
So a concrete prediction of our model would be that at very
low temperatures, the conductance corrections will not be flat
and moreover, there could be resonances in the transmission
due to the two barriers being seen coherently. A study of the
experimental graphs of Reilly {\it et al} \cite{reilly} leads us to speculate
that perhaps 
such resonances have already been seen.

\section{Model in a magnetic field}

In this section, we will study the effects of an in-plane magnetic field on 
the conductivity of a quantum wire. 
Since, orbital motion is not possible in an in-plane magnetic field,
we will only consider the 
effect of the Zeeman term. This term couples differently to spin up 
and spin down electrons; here up and down are defined with respect to the 
direction of the magnetic field which may or not be parallel to the quantum 
wire. Thus the $SU(2)$ symmetry of rotations is explicitly broken and 
the spin and charge degrees of freedom do not decouple any longer. 

For
very strong magnetic fields ($\gg 16 T$)
(Zeeman energies $\gg$ inter-subband energies),
all spins are polarised and 
conductances are quantised as  $e^2/h$ and the
wire  is essentially described as
a spinless LL.

But when the magnetic field is not that  large, but still large 
($8T\ll b\ll 16 T$),
each sub-band splits into two
 - a spin $\ua$ sub-band and
spin $\da$ sub-band.
We might expect barrier renormalisations for both 
the sub-bands to be the same.  But it turns out that the
barrier renormalisation for 
spin parallel (moments aligned with the magnetic field) band
is much smaller 
than that for the spin anti-parallel (or anti-aligned moments) band.
Hence, the conductance correction is much more for even
bands and negligible for odd bands. 

In fact, this odd-even effect, even exists when the magnetic field
is not large enough to completely spin-split the bands. Hence, one
would expect that the electrons with moments aligned with the magnetic
field, go through the quantum wire, more easily than those anti-aligned.

This odd-even effect, in fact, has been experimentally 
seen \cite{liang},
and perhaps could be instrumental in constructing a spin-valve using
these wires\cite{us2}.

\section{Summary and future prospects}

In this paper, we have introduced a non-homogeneous 
LL model for  the quantum wire,
motivated by the experimental results. 
The main features of the experiments that we have explained
using our model are

\bi
\item{} Flat plateau renormalisation\\
This was essentially done by  
modeling the  contact regions as LLs with  barriers between the contacts
and the leads and by keeping the  
contacts independent of gate voltage. The results were found to be
in qualitative agreement with experiments.

The main point to note here is that, in the absence of scatterers in the
wire,   
conductances are only dependent on the 
contacts. The 
internal wire region can  only be experienced if it has scatterers.

\item{} Odd-even effect in presence of magnetic field \\
We explained the odd-even effect in the 
presence of an in-plane magnetic field, by showing that  
the barrier renormalisation due to
interactions is different 
for $\uparrow$ and $\downarrow$ bands,
in the presence of the magnetic field.

\ei

By explicitly fitting a power law form to the conductance corrections as 
a function of temperature as given in the inset of Fig. 3  
in Ref.\cite{yacoby}, we find that 
\beq
\delta g = - 0.3512 ~T^{-0.1058 - 0.0345 T}~. 
\eeq
In the future, we would like to understand the reasons for the 
discrepancy of this law from the simpler power law form 
that our model predicts 
and perhaps make a more
realistic model, which could agree with the data more quantitatively.
We would also like to understand several features which are observed
on the rise between two successive plateaus, 
such as the ``$0.7$ effect''. For this,  one needs to study the 
model when some sub-band is partially opened.

\end{document}